\documentclass[acmsmall]{acmart}
\AtBeginDocument{%
  \providecommand\BibTeX{{%
    \normalfont B\kern-0.5em{\scshape i\kern-0.25em b}\kern-0.8em\TeX}}}

\setcopyright{acmcopyright}
\copyrightyear{0000}
\acmYear{0000}
\acmDOI{00.0000/0000000.000000}

\acmJournal{JACM}
\acmVolume{00}
\acmNumber{0}
\acmArticle{000}
\acmMonth{0}

\begin{document}

\title{TWIST-GAN: Towards Wavelet Transform and Transferred GAN for Spatio-Temporal Single Image Super Resolution}

\author{Fayaz Ali Dharejo}
\authornote{Both authors contributed equally to this research.}
\email{fayazdharejo@cnic.cn}
\orcid{0000-0001-7685-391}
\author{Farah Deeba}
\authornotemark[1]
\email{deeba@cnic.cn}
\orcid{0000-0002-1987-2402}
\author{Yuanchun Zhou}
\authornote{are corresponding authors.}
\email{zyc@cnic.cn}
\orcid{0000-0003-2144-1131}
\affiliation{%
  \institution{Computer Network Information Center, Chinese Academy of Sciences, University of Chinese Academy of Sciences}
  \city{Bejing}
  \country{China}
  \postcode{100190}
}

\author{Bhagwan Das}
\affiliation{%
  \institution{Department of Electronic Engineering, Quaid-e-Awam University Engineering Science and Technology}
  \city{Nawabshah}
  \country{Pakistan}}
\email{engr.bhagwandas@hotmail.com }

\author{Munsif Ali Jatoi}
\affiliation{%
  \institution{Department of Biomedical Engineering, Salim Habib University}
  \city{Karachi}
  \country{Pakistan}
}

\author{Muhammad Zawish}
\affiliation{%
 \institution{Telecommunication Software and Systems Group, Waterford Institute of Technology}
 \streetaddress{Rono-Hills}
 \city{Waterford}
 \country{Ireland}}

\author{Yi Du}
\authornotemark[2]
\email{duyi@cnic.cn}
\orcid{0000-0003-3121-8937}
\author{Xuezhi Wang}
\email{wxz@cnic.cn}
\orcid{0000-0001-5222-248X}
\affiliation{%
  \institution{Computer Network Information Center, Chinese Academy of Sciences, University of Chinese Academy of Sciences}
  \city{Bejing}
  \country{China}
  \postcode{100190}
  }

\renewcommand{\shortauthors}{Fayaz and Farah, et al.}

\begin{abstract}
  Single Image Super-resolution (SISR) produces high-resolution images with fine spatial resolutions from a remotely sensed image with low spatial resolution. Recently, deep learning and generative adversarial networks (GANs) have made breakthroughs for the challenging task of single image super-resolution (SISR). However, the generated image still suffers from undesirable artifacts such as, the absence of  texture-feature  representation  and  high-frequency information.  We propose a frequency  domain-based spatio-temporal  remote sensing single image super-resolution technique to reconstruct the HR image combined with generative adversarial networks (GANs) on various frequency bands (TWIST-GAN). We have introduced a new method incorporating Wavelet Transform (WT) characteristics and transferred  generative  adversarial  network.  The  LR  image  has  been  split  into  various  frequency  bands  by  using  the  WT,  whereas,  the transfer generative adversarial network predicts high-frequency components via a proposed architecture. Finally, the inverse transfer of wavelets produces a reconstructed image with super-resolution. The model is first trained on an external DIV2 K dataset and validated with the UC Merceed Landsat remote sensing dataset and Set14 with each image size of 256x256. Following that, transferred GANs are used to process spatio-temporal remote sensing images in order to minimize computation cost differences and improve texture information. The findings are compared qualitatively and qualitatively with the current state-of-art approaches. In addition, we saved about 43\% of the GPU memory during training and accelerated the execution of our simplified version by eliminating batch normalization layers.
\end{abstract}



\maketitle

\section{Introduction}
Notably, there are numerous archived fine spatial resolution remotely sensed images accessible the worldwide. These archived fine spatial resolution datasets may provide a valuable prior land cover pattern, resulting in the spatio-temporal mapping. Since many HR images can be generated within one LR image, the Single Image Super-Resolution (SISR) is considered an ill-posed problem. Although SISR has several advancements, the question remains: how can photo realistic effects be retrieved by more natural textures and less noisy objects. To that end, new approaches and learning strategies are  being  introduced  successively  \cite{hanif01}. It  should  be  noted  that  a  learning  method  generates  an  HR  image  by  learning  a nonlinear  mapping  from  LR-to-HR  through  a  deep  neural  network. In  this  approach,  we  chose  to expand  the  intended method by using the advanced method Generative Adversarial Network for image super-resolution (SRGAN).

\begin{figure*}[ht]
\includegraphics[width=1\textwidth]{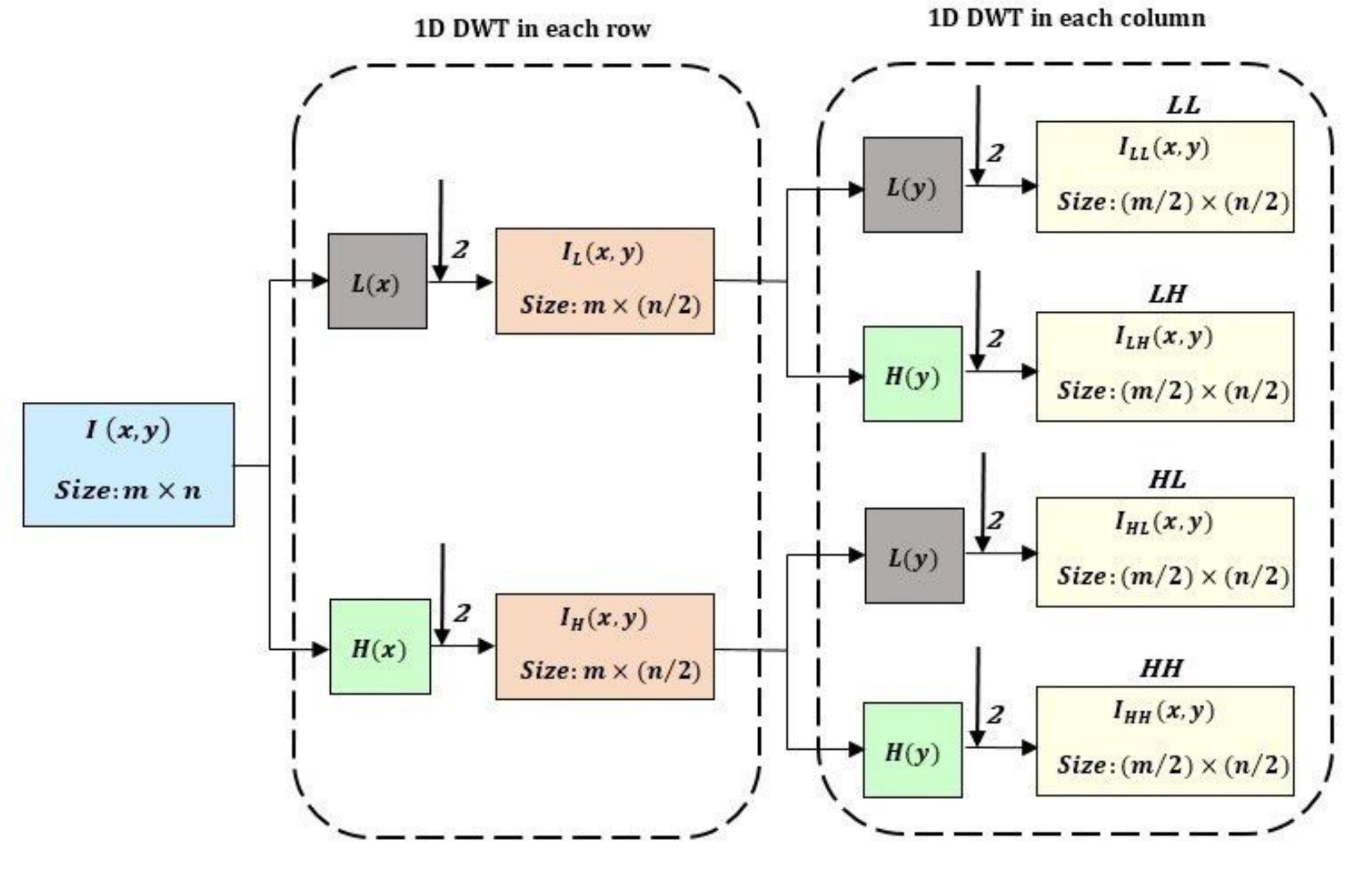}
\caption{2D-DWT single-level for I, which can be split into two stages. First, on each row of I, $(m\times n)$ is implemented and second intermediate subbands $[m \times (m/n)]$gain 1-DWT is implemented into two intermediate subbands.}
\centering\label{Fig1}
\end{figure*}

Spatio-temporal data collection is a rapidly evolving field in which powerful computer processors (GPUs) for large-data analysis  are used.  Space-time  databases  contain  data  collected  in a  given  place  and  time  over  both space  and  time  that define an occurrence. A variety of spatio-temporal applications with rich details needs to convert LR images to HR images, such as target detection \cite{naqvi01}, recognition, and other high-resolution images \cite{deeba01}. Incorporated by a technology called super-resolution  (SR),  a  number  of  researchers  chose  to  reconstruct  high-resolution  (HR)  images  from  low-resolution  (LR) images instead of focusing on physical imaging \cite{deeba02}. Ultimate spatio-temporal objects would be preserved and handled by a spatial database management system (SDBMS), which provides spatial capacities, including spatial data and operations models.  The  artifacts  of  space  or  time  are  not  limited  to  2D  or  3D geometrical data.  Spatio-temporal  deep  learning  is becoming widely attractive and is essential to increase the resolution to retrieve the information from distinct applications such as GPS devices, internet-based map services, weather services, digital Earth, and satellite. The SISR is used to restore the high-resolution (HR) image in relation to the low-resolution (LR) issue \cite{deeba02} \cite{irani01} \cite{deeba03}, given that the information on high frequencies must include HR images. Super-resolution can be commonly used in several applications, like medical imaging \cite{shi01},  satellite  imaging \cite{mat01} \cite{dharejo01},  and  health,  safety,  and  monitoring \cite{zou01}\cite{nazir01},  where  high-frequency  requirements  are  quite relevant. SISR is a skewed problem, as it is appropriate to expect more HR pixels than the corresponding LR image \cite{yang01}. The example-based approach has been adapted to use this additional information by current methods, which either explore the self-similarities \cite{freed01}\cite{yang02}, or map the LR counterpart in HR patches using external samples \cite{kim01}\cite{chang01}\cite{tim01}.

Further,  to  use  sparse  coding  before  assuming  that  an  image  might  be  a  well-designed  dictionary,  the  image  will  be  a reasonable  reference  and  is  commonly  used  for  sparse-code  methods  \cite{lu01}\cite{lu02}.  With  the  rapid  growth  of  deep  learning theory in recent years, deep neural networks have been widely used in various spatio-temporal remote sensing tasks; such as the remote sensing images scene-classification \cite{lu03}, hyperspectral target image detection \cite{lu04}, image classification \cite{lu05}, and in multi spectral change detection \cite{zhang01}. Due to their good learning ability, DL-based methods have also been used to solve  SR-based  inverse  problems,  and  the  discovery  of  non-DL-based  SR  methods  has  been  enhanced \cite{glas01}\cite{ma01}\cite{dharejo02}\cite{tim02}.  To significantly predict LR and HR image pairs nonlinear mapping, Dong et al. \cite{dong01} uses CNNs and beats the non-DL-based SR methods.

\begin{figure*}[ht]
\includegraphics[width=1\textwidth]{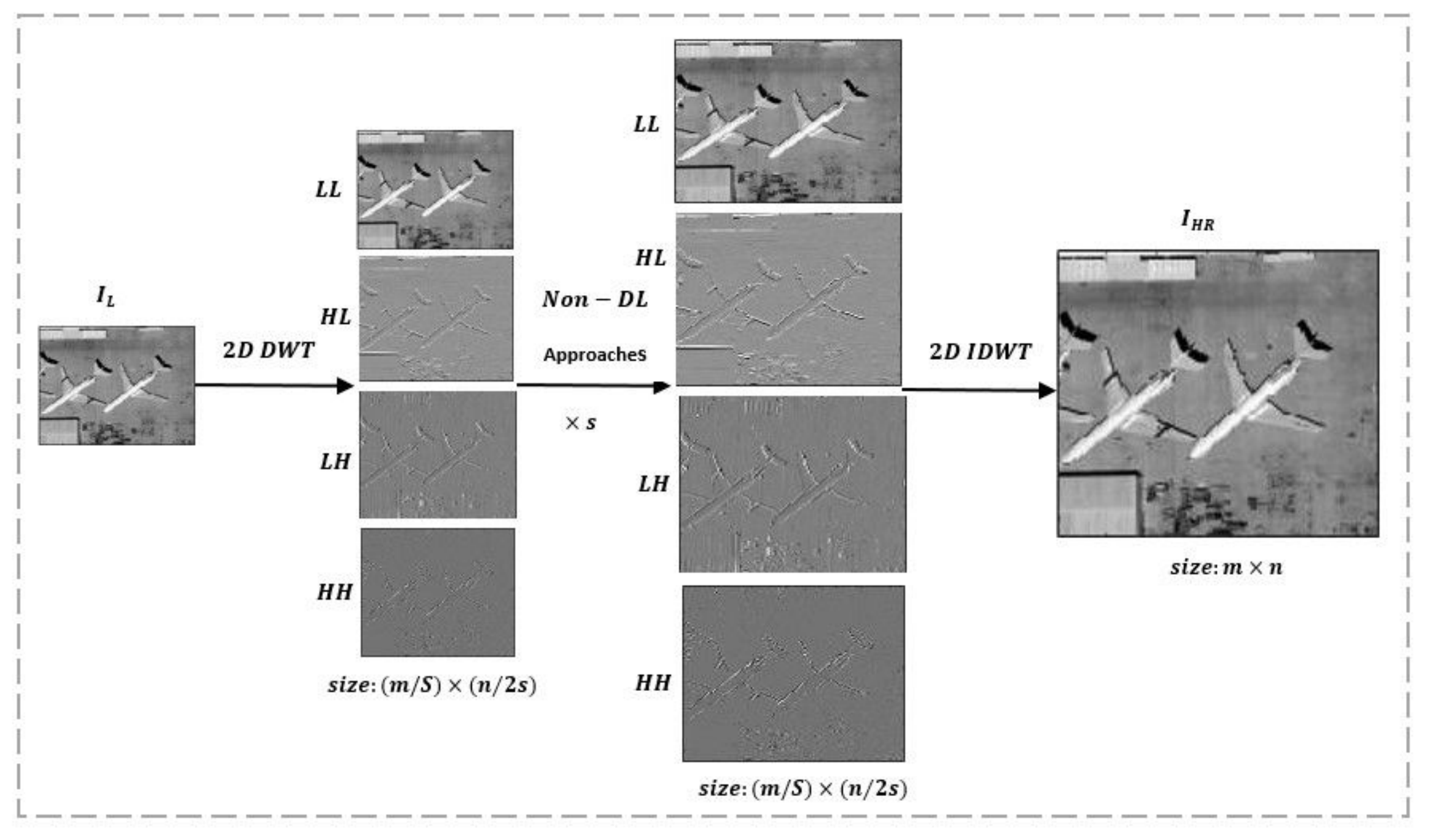}
\caption{Flow chart of the method based on non-DL SR DWT. First, through the non-DL SR method, we decompose the LR image $I_{LR}$ into four wavelet components, and then increase the scale. Finally, combine the inverse 2-DWT operation as a super-resolution subband into the reconstructed HR image $I_{HR}$}
\centering\label{Fig2}
\end{figure*}

Shi et al.\cite{shi02} proposed a powerful CNN sub-pixel (ESPCN) that can minimize runtime by providing the LR image as input and preparing the network input HR image function map (rather than interpolating the LR version image). Dong et al. \cite{dong02} proposed an accelerated lightweight network structure known as Super-Resolution Convolutional Neural Network (SRCNN).  Kim  et  al. \cite{kim02}  have shown a  profound super-resolution  network  based  on  the  achievement  of  highly  deep networks \cite{he01} trained on ImageNet \cite{kriz01}. They have used multiple HR and LR images with upgrade factors to minimize computation  complexity  speed  up  training.  Kim  et  al.  \cite{kim03}  introduced  more  convolution  layers  to  avoid  adding  a  new parameter as the network depth was increased, and implemented the Deep Recurring Neural Network [DRCN]. Later the Deep Residual Network (DRRN) was proposed by Tai et al. \cite{tai01}, which attempts through residual learning, to build large but concise networks. As for the spatio-temporal remote sensing image resolution concerned, by using a generative model, Haut et al. \cite{haut01} recognize image distribution and suggested a supervised method of SISR. Local and global residual learning \cite{lei01} is incorporated into the reconstruction of supervised remote sensing images. Reconstruction with super-resolution can also  help  to  distinguish  hyperspectral  images  \cite{hao01}. Most  existing  SISR  DL-based  methods  perform reconstructing  the spatial domain and aim to increase LR pixels resolution and their HR image equivalents. In DL-based SISR strategies, this technique has been neglected, although it seems easier to restore missing high-frequency data within the frequency domain. Figure. \ref{Fig1} shows the single-level 2D-DWT method for $\boldsymbol{I}$, resulting in four subbands.

The wavelets have unique features, like sparse wavelet subbands; therefore, multi-scale  modeling helps represent the image in better quality. Wavelet transform (WT) can extract information and perform multi-resolution analysis, often used in signal processing \cite{aba01}\cite{mallat01}\cite{abbate01}. The shift in wavelets has also been a very significant and efficient way for a multi-resolution image to be represented and stored \cite{mallat02}. It displays the spatial and textual details of the image on different levels. A wavelet method  for  compressing  remote  sensing  images \cite{tian01},  a  flexible  Galerkin-Wavelet  approach  for  restoring  objects  at  a limited-angle tomography, is included in several remote sensing transformations \cite{lu06}. he  wavelet  transformation  was  widely  adopted  to  support  spatio-temporal  imagery  spatial  resolution \cite{chan01}\cite{kin01}\cite{kim04}\cite{li01}.The combination of wavelet transform and interpolation algorithm completes SR spatio-temporal remote sensing image task \cite{tao01}. Spatio-temporal  remote  sensing  images  data  is  acquired  using  our  method  from  a  multi-temporal  and  multi-viewpoint dataset,  which  can  be  considered  a  type  of  spatio-temporal  remote  sensing  image.  Non-redundant  information  can  be obtained from a spatio-temporal remote-sensing image, enhancing useful information in the spatial domain and improving texture-feature representation. For this reason, it is an effective method for super-resolution reconstruction that utilizes the important information given by the spatio-temporal remote sensing image. As a result, the performance with which remote-sensing image data is used can be realistically enhanced. In  this  paper,  a  three-level  super-resolution  remote  sensing  image  technique  is  proposed  by  combining  transform wavelet  and  transferred  generative  adversarial  networks  detail  enhancement  based  on  spatio-temporal  remote-sensing images. The first part incorporates the division in four subbands of the LR image through a 2-D discrete wavelet transform (DWT)  and  replacing  the  LR  image  with  the  low-frequency  subband.  Together  the  four  subbands  are  inserted  into  the proposed network, as shown in Figure \ref{Fig4}. Our proposed transmitted adversary network goal is to predict the residually improved  HR  wavelet  component  of  the  four  corresponding  sub-bands  of  single  image  super-resolution  using  basic generative adversarial network (SRGAN) architecture\cite{ledig01}. We mainly used the Transferred GAN to remove the redundant layers to decrease the memory burden and speed up the network. We train our model with an external data set DIV2K \cite{tim03} from previous knowledge of transfer learning and then fin-tune the network with remote sensing image datasets. Finally, by using the inverse 2D-IDWT transform, the HR reconstruction images are obtained.  It is noticed that the network has become simpler by eliminating the redundant layers, and the computational network speed is much improved. Our paper has four main key contributions:

\begin{enumerate}
  \item The  proposed  method is  spatio-temporal  remote  sensing  frequency-domain  SISR  to  take  full  advantage  of displaying images at different frequency bands (wavelet sub-bands).
  \item The  batch  normalization  layers  are  removed  to  minimize  memory  usage,  processing  burden,  and  enhance accuracy, unlike previous GAN-based SR approaches.
  \item Our model is designed to cope with inadequate training in a transfer-learning way to implement deep learning techniques for spatio-temporal remote-sensing applications. 
  \item To  further  improve  the  performance,  the low-frequency  wavelet  portion  is  replaced  with  a  more  accurate  LR image. The results indicate that the proposed approach outperforms most state-of-the-art techniques to validate objective and subjective assessment accuracy. 
\end{enumerate}

The continuation of our paper is structured as Wavelet Transform (WT) details are in session II. The proposed approach is defined  in  session  III.  The  experimental  setup,  qualitative,  and  quantitative  analysis  are  in  session  IV,  and  they  finally concluded in session V.

\section{Background}

We  process  the obtained spatio-temporal  remote-sensing  images  in  order  to  fully  exploit  complementary  information between spatio-temporal remote-sensing images and create SR reconstruction images containing much detailed textures. We will briefly present the Wavelet Transform (WT) theory in this section and the conventional SISR approach based on DWT.

\subsection{An Overview of Wavelet Transform}
When the signal is decomposed in an ascending hierarchy, the wavelet transform collects time and frequency information \cite{li02}.
The stable 1-D WT (1D-CWT) of the signal x(c) is given in equation (1),

\begin{equation}
F(a,b)= \int_{-\infty}^{\infty} x(c) \psi_{a,b} (c) dc
\end{equation}
Where $\psi_a,_b(c)$ is the version of the  mother wavelet dilated and translated as,
environment.  

\begin{equation}
\psi_a,_b (c)=\frac{1}{\sqrt{a}}\psi{\left(\frac{c-a}{b}\right)}
\end{equation}
Where a and b are parameters for dilation regulation that regulate the Wavelet translation dilation.
DWT is more proper to use to analyze discrete signals \cite{aha01} instead of using the CWT. The $\leq x[d]$ discrete signal is set to 1D-DWT

\begin{equation}
   G(k,l)= \sum_{d=-\infty}^{\infty}x[d]\psi k, l(d)
\end{equation}
Where $\psi k,l (d)$ the mother wavelet is dilated and translated version and can be measured as,

\begin{equation}
   \psi k,l (d)=2^\frac{-k }{2} \psi[2^{-k} d-l]
\end{equation}
In reality, DWT would be used to transmit the input signal through the low-pass filter L(e) and high-pass filters H(e). Then to decrease its approximation and the accuracy of the input signal by 2 \cite{ledig01}. The Haar wavelets are defined as, L(e)  and H(e),
\begin{equation}
  L(e)=
    	\begin{cases}
    	    1, & e=0,1 \\
            0,  &  otherwise\\
        \end{cases}
          H(e)=  
         \begin{cases}
    	 1,  &   e= 0 \\
        -1,  &   e= 0\\
         0,  &   otherwise\\
        \end{cases}
\end{equation}

For the 2-D image I, the pixel value of the xth and yth column should be indicated by I(x, y). 2D-DWT I is independently considered in each dimension accordingly; in columns and row sites, another term is used 1D-DWT. The 2D-DWT is divided into four substrates representing the average components; \textbf{LL},\textbf{LR}, \textbf{HL}, \textbf{HH}. 
 
\subsection{SR Based on the Discrete Wavelet Transform (DWT)}

The ability to properly reconstruct information is limited to interpolation-based SR methods. Since high-frequency domain information cannot be recovered well during SR, the result is not exact. Edges must be kept to increase super-resolved image performance. The DWT was used to maintain the image elements high frequency; for some of its high-frequency components, the $\boldsymbol{LL}$ portion with DWT decomposition, and the$\boldsymbol{LR}$, $ \boldsymbol{HL}$ , and $ \boldsymbol{HH}$  components are also structural data.
DWT is combined with Non-DL SR methods by conventional DWT-based SR methods. 
    
Let $I_G$ indicate the image of grounds truth HR in $m \times n$ by, and $I_L$ denote LR in size $\left(\frac{m}{s} \right)\times \left(\frac{n}{s} \right))$ by the same size (s), which is a scale factor. Allow the $\boldsymbol{I_{LR}}$  reflect LR up-scale performance, with m×n  size, via bicubic interpolation, and $\boldsymbol{I_{HR}}$ to display the HR image reconstructed, with $\boldsymbol{m \times n}$ size.

\begin{enumerate}
  \item Figure. \ref{Fig2} illustrates the DWT non-DL SR method's flowchart, consisting of three steps; 2D-DWT, i.e., LL, HL, LH, and HH, decomposes IL into four separate subbands.
  \item For each of the four subbands, a particular non-DL SR method is adopted for increasing their spatial resolution in s times.
  \item The HR image of IHR is reconstructed, employing a discrete reverse WT (IDWT).

\end{enumerate}

\subsection{Generative Adversarial Networks}
\begin{figure}[ht]
\includegraphics[width=1\textwidth]{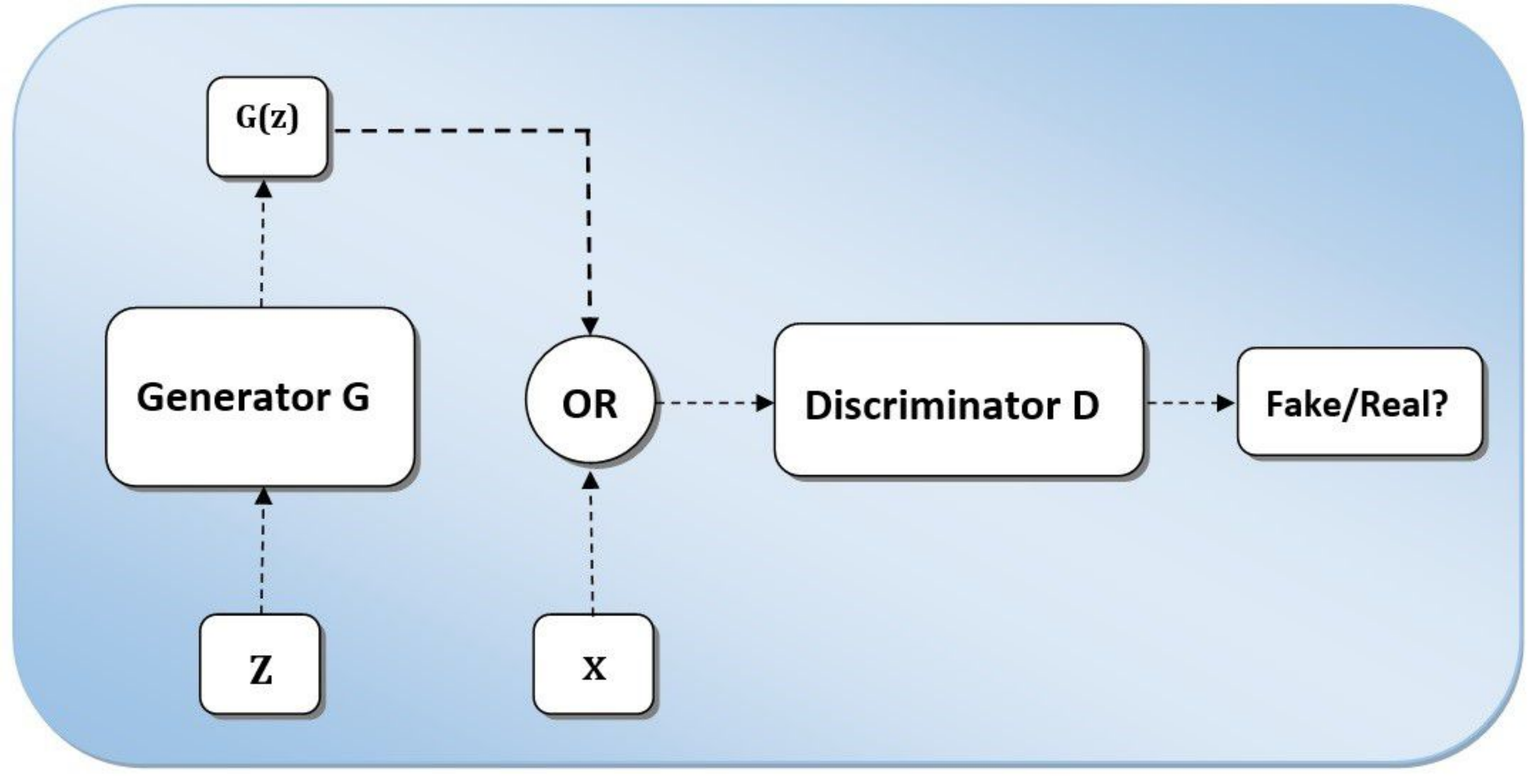}
\caption{The general architecture of GAN.}
\centering\label{Fig3}
\end{figure}

Generator G, which transforms a sample from a random uniform distribution to a given data, discriminator D, which calculates the probability of a sample being distributed by a data sample; \cite{dai01} (GANs) is a two-characteristic class of data dispersion  models.  The  game-theoretical  min-max  concepts  are  the  origin  of  the  generator  and  discriminator.  They  are usually  taught  in  tandem  by  combining  D  and  G  instruction.  Although  GANs  can  visually  attract  images  with  high-frequency  details,  GANs  still  face  many  unanswered  challenges,  which  means  they are  challenging  to  train.  Recently, researchers have analyzed different aspects of GANs, such as the use of an accurate variable \cite{mirza01}, training optimization \cite{saliman01}, and the use of task-specific cost functions \cite{cres01}. In addition, Shahin et al. \cite{mahdi01} explore an alternative viewpoint on discrimination, which is different from the traditional view of probabilism as the paradigm of discrimination. In Figure \ref{Fig3}, the most generic GAN architecture is given.

\section{Methodology}
\begin{figure*}[ht]
\includegraphics[width=1\textwidth]{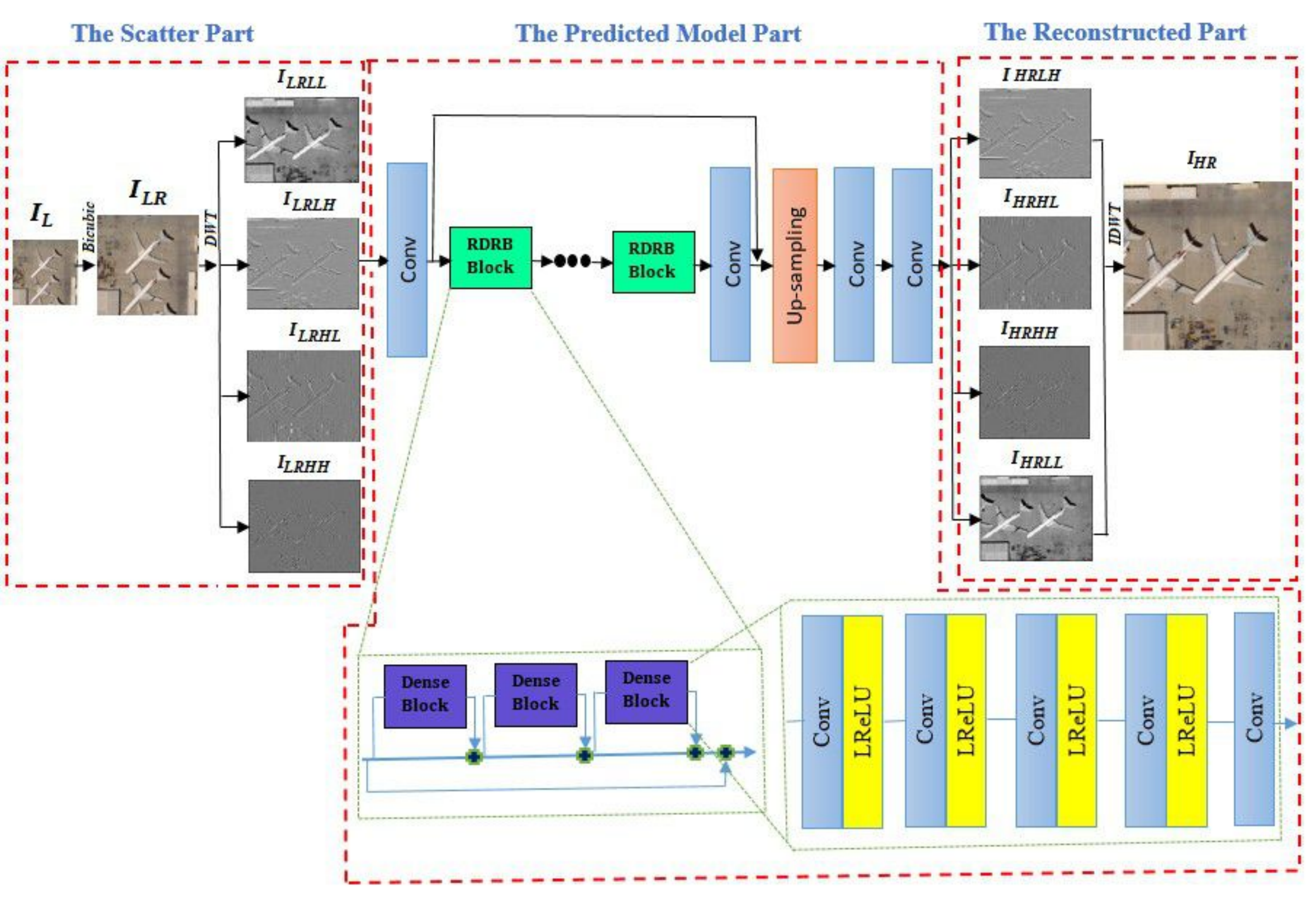}
\caption{The method flow chart proposed by TWIST-GAN includes three parts; decompose parts, predict elements and reconstruct parts. The decomposed part decomposes the interpolated version of the LR image into four components of the LR wavelet. The prediction part uses a designed GAN-based SR network to estimate the LR counterpart's HR wavelet component. 2D-IDWT is used to generate super-resolution images based on the HR wavelet components of the reconstructed part. The residual blocks are substituted with RRDB blocks. It is worth noting that RDDB contains dense blocks with ReLU and no BN layer.}
\centering\label{Fig4}
\end{figure*}

Once the corresponding wavelet elements are accurately predicted, high-quality HR images can be recovered from the LR image with comprehensive texture details and global topology information. Compared to CNN models, GANs have improved their performance and thus increased the quality of the image. GANs can enhance the consistency, appearance, and  colour  of  images,  build  faces,  and  perform  many  more  interesting  tasks;  GAN  achieves  good  extraction  and representation capacity. It thus can transform the reconstruction process of an HR image into the prediction of its wavelet portion.  The two above facts allow us to put the TWIST-GAN schemes for the super-resolution challenge together with a GAN network and WT. The Figure.\ref{Fig4} shows the entire architecture of Spatio-Temporal Remote Sensing Super-Resolution Combined with the Transferred Generative Adversarial Network and Wavelet Transformation (TWIST-GAN). It can be divided as the part disassembled, the part predicted, and the part reconstructed.

\subsection{Wavelet-based Scatter Part}
Wavelet has various numerical methods that allow valuable information to be extracted from the signal and signals to be evaluated with different resolutions. Wavelet is an efficient multi-resolution solution for different imaging resolutions, such as high frequency and low-frequency images \cite{lu03}\cite{lu04}\cite{lu05}\cite{zhang01}. The image is separated into independent details by means of a wavelet  transformation.  The  wavelet  coefficients  can  be  interpreted  as  vectors  for  the  image.  The  high-frequency components can mirror the differences in the frequency in different directions to increase the texture detail. We train image coefficient wavelets to achieve the objective of reconstruction of remote sensing image super-resolution. Discrete packet wavelet decomposition and two-dimensional image reconstruction often used Haar wavelet as a wavelet base feature. It has a simple function, and the application is growing, particularly the reinforced symmetry that will allow us to avoid phase distortion during image decomposition.

We  first  obtain  the ILR  from  the  LR  image IL  for  the  first  part  that  is  disassembled.  We  decompose  by  using  Haar wavelet decomposition into four wavelet components ILRLL, ILRLH, ILRHL, and ILRHH. From Figure. \ref{Fig2}, Compared to IL, it can be seen that ILRLL has less detail, so we replaced the ILRLL component with IL, which is an input to the network that further helped boost the super-resolution quality. The disassembled part can be summarized as:
\begin{itemize}
   \item Zoomed-in IL to obtain ILR using bicubic interpolation with a scale factor of ×s
   \item By sung 2D-DWT decomposition, ILR can be decomposed into  ILRLL, ILRLH, ILRHL, and ILRHH.
   \item Load   IL, ILRLH, ILRHL, and ILRHH into the GAN network that has been designed.
\end{itemize}

\subsection{Predicted Model Part}
\begin{figure*}[ht]
\includegraphics[width=1\textwidth]{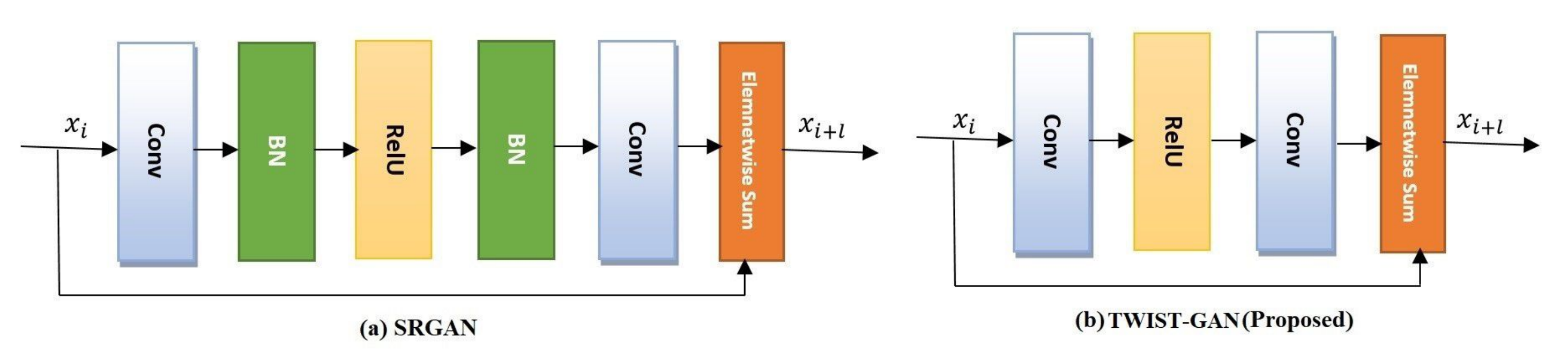}
\caption{Comparison of the block architecture. (a) SRGAN’s block architecture (b) The proposed method's block architecture.}
\centering\label{Fig5}
\end{figure*}

The model TWIST-GAN with the structure provided in Figure.\ref{Fig2} is used in the predicted portion to predict the wavelet components of the HR image from their LR counterparts, denoted by $\boldsymbol{I_{HRLL}}, \boldsymbol{I_{HRLH}}, \boldsymbol{I_{HRHL}}, and \boldsymbol{I_{HRHH}}$. Alternatively,  the  SRGAN  and  proposed  architectures  are  shown  in  Figure. \ref{Fig5}(a,b),  respectively.  The  batch normalization layers are omitted, providing more apparent evidence that the residual component batch normalization layers in the proposed architecture have been omitted. As indicated in EDSR \cite{dharejo02}, batch normalization (BN) layers are removed to decrease memory utilization and computational complexity. The entire TWIST-GAN network architecture is shown in Figure.\ref{Fig4}.

In  the  same  framework,  we  overlay  a  residual  network  in  the generative network.The  residual  network's  core component is that the deep network can effectively control the generative model and reduce GAN preparation complexity. Recent  research  has  developed  residual  high-performance  networks  for  low  and  high-speed  computer  vision  tasks. Although SRRes Net can be successfully implemented and deliver excellent efficiency in SRGAN, we  work to simplify the  network  even  further,  although  keeping  good  performance.  To  normalize  the  characteristics  and  reduce  the heterogeneity  \cite{lim01},  Nah  et  al. \cite{nah01}  have  implemented  the  elimination  of  batch  normalization  layers  from  the  network. However, the Batch Normalization (BN) layers are useful for the classification task, but the BN layers are not appropriate for the super-resolution task, as shown in Figure.\ref{Fig5}. As our strategic growth, the simplified network can display better image resolution performance than the original version. In addition, the batch normalization layers cost a lot of GPU memory and slow down the device rate. We can somewhat save about 43\% of GPU memory during the workout and speed up running with our simplified version.

\subsection{The Reconstruction Part}
To reconstruct the HR super-resolved image IHR, we use a 2D-IDWT inverse wavelet onto four components $\boldsymbol{IHRLL}, \boldsymbol{IHRLH}, \boldsymbol{IHRHL}, and \boldsymbol{IHRHH}.$

\subsection{Network Architecture}
We proposed TWIST-GAN by inspiring model SRGAN \cite{ledig01}, the configuration of the discriminator D is the same in our method as in SRGAN, but the G generator is changed to boost the SR outcome, as shown below: 
\begin{enumerate}
  \item The residual block is replaced by the residual-in-residual block called RDDB to use dense connections and multi-layer residual networks \cite{wang01}
  \item We have found from SRGAN that GAN suffers from training instability, so the original GAN is replaced with WGAN \cite{gul01} to resolve this deficiency. We are complementing G planning and helping to produce more realistic results.
  \item From EDSR \cite{lim01}, we found that eliminating the layers of batch normalization can improve the network in terms of results and computation complexity.

\end{enumerate}

RDDB can be defined as residual learning applied at different points to form a residual-in-residual (RDDB) structure. Short distances between a layer and each layer are built within each dense block; the block's data flow could be linked. We use 25 RRBD blocks in our context. In addition to improving G architecture, BN layers are excluded. Removal of BN layers will bring several benefits. To eliminate the network complexity, BN-layers would first standardize the functionality. Secondly, BN layers consume extensive memory processing and GPU commonly. This can improve the network's performance, particularly with limited computational resources, by removing batch normalization layers. When the network is very deep, and with an adversary's learning  \cite{wang01},  unnecessary  artifacts  may  be  added  because of  BN  levels.  For  the  above  factors,  the  BN  layer  has  been eliminated.

\section{Experiment}
\subsection{Train the network via Transfer learning}
\begin{figure*}[ht]
\includegraphics[width=1\textwidth]{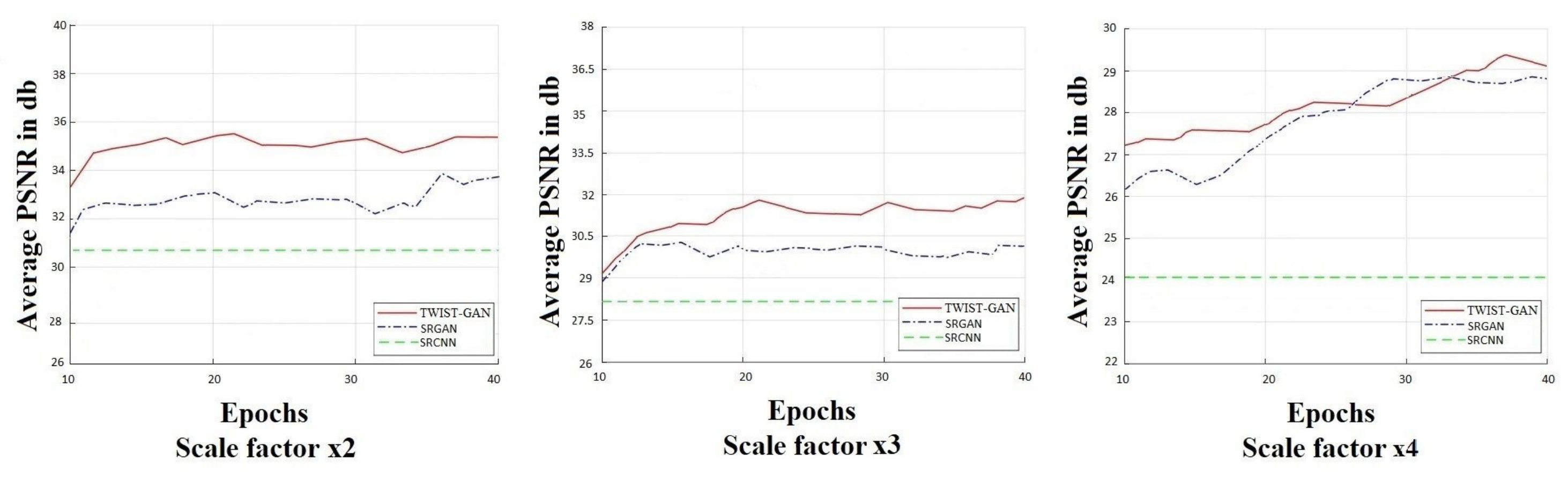}
\caption{Concerning the progressively increased network depth, the output curves of the proposed TWIST-GAN on the UC Merceed dataset. As the test baselines, SRCNN and SRGAN are used. Note that SRCNN’s output will decrease with the increase in its depth, but we purposely fix the results to their corresponding peak values for better comparison. For SRGAN, for the sake of fairness, the experimental findings recorded in Figure 3are used directly.}
\centering\label{Fig6}
\end{figure*}

The super-resolution of deep-learning spatio-temporal remote sensing images has more problems than the natural one. The research ideas of sufficiently strong samples are the foundation of deep learning. However, many high-quality satellite images that meet these criteria are not easy to obtain. Consequently, knowledge transfer from external data sets has received tremendous interest due to the continuous deep learning establishment. Transfer learning focuses on solving various problems in conjunction with areas using shared information, which leads to  addressing  tasks  in  an  area  based  on  experience  gained  from  other  environments.  If  only  a  few  training  samples  are available, we may apply external knowledge through translation learning to the target domain.An external DIV2K dataset \cite{tim03} is used to prepare the proposed model in this paper. DIV2K is a contemporary image retrieval repository. There are 1,000 natural objects, all between 2000 and 1400 pixels, with a reasonable resolution. We prepare  for  the  proposed  model,  especially,  provide  qualified  training,  and  optimize  the  network.  Participating  images chosen from the objective datasets for the DIV2k training boost the per-trained network \cite{tim03}. The pre-training network is optimized by obtaining images for training from the UC Merced dataset, a typical dataset of 256x256 pixel remote sensing image classification scenes, as shown in Figure.\ref{Fig7}. UC Merced contains a total of 21 natural images, with 100 images of each  type.  Finally,  we  choose  the  order  targeting  airplane  and  80  airplane  images  as  targets  training  samples,  so  the remaining 20 are used for testing. Figure.\ref{Fig6} shows all the curves of performance grouped according to their respective test scales.

\subsection{Training Details}

Above all, we explain the meaning of the relevant symbols. Let $\boldsymbol{X_i}$ represents the image of $\boldsymbol{ith}$ ground-truth HR $X_i-$, $X_{i-LH}$,  $X_{i-HL}$, $X_{i-HH}$ denoting vector variants $\boldsymbol{X_i}$ wavelet components and $\hat{x}_i=[x_{i-}^T,X_{i-LH}^T,X_{i-HL}^T,
X_{i-HH}^T]^T$ denote how these components are concatenated. Similarly, for the reconstructed image, let $\boldsymbol{H_i} $ represents the $\boldsymbol{ith}$ reconstructed HR image. Where $x_i-, h_{i-LH},h_{i-HL},h_{i-HH}$ denoting vector variants of $H_i$ and  h $\hat{h}_i=[h_{i-}^T,h_{i-LH}^T,h_{i-HL}^T,
h_{i-HH}^T]^T$ denote how these components are concatenated.
 In the training process, bicubic interpolation on 800 HR (down-sampling factor 2, 3, and 4) images have been used to obtain the LR image. During the Sub-image scale 256x256 training, the HR image is randomly trimmed into shape. The network reconstructs the image by learning the mapping relation between the LR image's wavelet coefficients and the wavelet coefficients of the HR image. The network parameters are defined as the network learning rate is 0.0002, and the number of iterations is 28,000.\\
SISR task is to know a mapping function $ X=gw(Y)$ between both the LR image $\boldsymbol{Y}$ and the HR image $\boldsymbol{X}$, where the network parameter is defined by $ \boldsymbol{W}$. We measure the loss $\boldsymbol{l_2}$ of the wavelet components, different from other DL-based SR methods, the SR method measures the Euclidean distance between the real HR image and the reconstructed HR image as a loss. The technique to restore the real image and HR is as follows.

\begin{equation}
    loss=\frac{1}{N_B} \sum_{i=-1}^{N_B} \|\hat{x}_i-\hat{h}_i\|_2^2
\end{equation}
Where $N_B$ refers to batch size. The proposed approach describes its overall objective function as,

\begin{equation}
    arg_W ^{min}=\sum_{i=-1}^{N_B}\|\hat{x}_i-\hat{h}_i\|_2^2+\lambda\|\boldsymbol{W}\|_2^2
\end{equation}
The first terminology is fidelity, meaning that the HR wavelet's restored portion approximates the ground truth's HR portion \cite{lu01}. The second term is a regularization term to prevent overfitting, and the coefficient $\lambda$ is used to simplify the problem to simulate two terms together.    
  The proposed approach is implemented in TensorFlow using a single 8-GB NVIDIA GTX1080 GPU, costing training five days and a day in relation and fine-tuning separately. We understand that our approach takes more time than SRCNN \cite{dong02}, VDSR \cite{kim01}, and SRGAN \cite{ledig01}, DRCN \cite{kim02}, respectively, for training. This is because the network we have designed is more complicated than the networks used in comparative approaches. However, the proposed method could yield better outcomes than the approaches compared.

\begin{equation}
    M(X, H)=\dfrac{1}{mn}\sum_{p=1}^m\sum_{q=1}^n (X(p,q)-H(p,q))^2
\end{equation}

\begin{equation}
    PSNR(X,H)=10lg\frac{255^2}{MSE(X,H)}
\end{equation}
Where m and n correspond to the image's width and height, the larger the PSNR, the better quality image.
Structure Similarity Index: To evaluate the structural similarity between the ground truth HR and the reconstructed HR image H, the Structural Similarity Index (SSIM) \cite{nah01} is used, and defined as follows:
\begin{equation}
    SSIM(X,H)=l(X,H) c(X,H) s(X,H) 
\end{equation}
   
\begin{equation}
    l(X,H)=\frac{2\mu X\mu H+C_1}{\mu_X^2+\mu_H^2+C_1},c(X,H)=\frac{2\sigma X\sigma H+C_2}{\sigma_X^2+\sigma_H^2+C_2},s(X,H)=\frac{\sigma X\sigma H+C_3}{\sigma X \sigma H+C_3}
\end{equation}
Where $\mu X$ and $\mu H$ express the mean value of  $X$ and  $H$, respectively, $\sigma X$ and $\sigma H$ show the standard deviation of $X$ and $H$, respectively, $\sigma XH$ is the covariance between $X$ and $H$, and $C_1$, $C_2$, and $C_3$ are constants. A higher SSIM value indicates superior quality.

\subsection{Experimental results and analysis}
We compared the research method with the SRGAN \cite{ledig01} algorithm’s performance. On the training set DIV2K, we reproduced the suggested algorithm. The results of the SRGAN algorithm differ significantly from the test results since the different training sets. To assess the accuracy of the restored images with precision. In our method, we used PSNR, SSIM, FSIM, and UIQ as test indexes.Peak Signal-to-Noise Ratio (PSNR): The peak signal-to-noise ratio (PSNR) \cite{hore01} is an image quality measure and relies on the mean square error (MSE) between ground truth and the reconstructed image that can be expressed as:

\begin{figure*}[ht]
\includegraphics[width=1\textwidth]{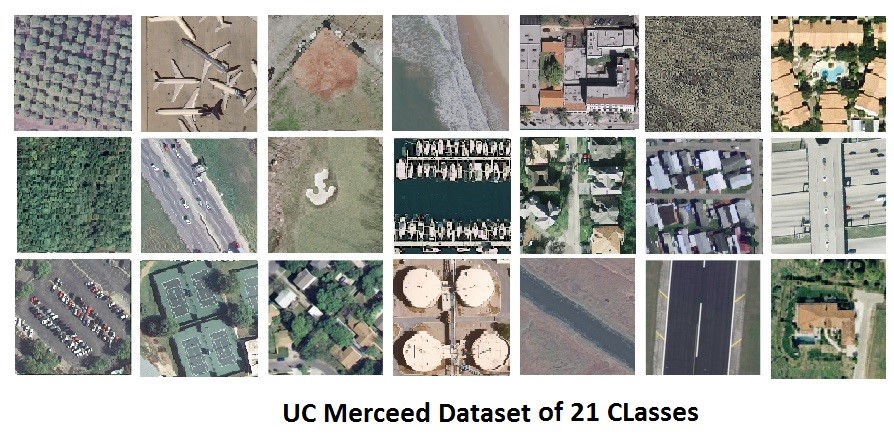}
\caption{UC Merceed dataset uses 21 classes to evaluate our approach, and each class has 100 images with 256x256 pixel images.}
\centering\label{Fig7}
\end{figure*}

\begin{table}\label{Tab1}
  \caption{The PSNR value of HR reconstructed image (db) compared to the latest methods}

  \begin{tabular}{ccccccccc}
   \hline
Image & Scale & Bicubic & SRCNN& FSRCNN & ESPCN& VDSR& SRGAN& \textbf{Proposed}\\
& & & \cite{dong01} & \cite{dong02} &\cite{shi02}&\cite{kim01}&\cite{ledig01}& \textbf{(TWIST-GAN)}\\
    \midrule
    airplane41 & x2 & 28.56 & 31.88 & 33.01 & 32.78 & 34.14 & 34.18 & \textbf{35.15}\\
& x3 & 25.66 & 28.13 & 29.47 & 29.19 & 30.27 & 30.36&\textbf{31.23}\\
& x4 & 24.12 & 25.97  & 26.21 & 25.22 & 27.24 & 27.45&\textbf{28.28}\\
 \hline
  airplane47 & x2 & 29.16 &30.89 &32.12&33.07&34.43 &34.33& \textbf{34 .61}\\
& x3 & 26.61 &27.93&28.67&29.90&30.67 &30.55&\textbf{32.03}\\
& x4 & 25.34 &24.91 & 25.71&26.45&27.88&27.76& \textbf{27.91}\\
 \hline
 airplane90 & x2 & 28.45 &31.66 &31.94&31.86&33.34&33.77 & \textbf{35.16}\\
& x3 & 25.89 &27.83 &29.06 &28.59&29.47&29.66&\textbf{29.86}\\
& x4 & 24.16 &25.32& 25.61&24.68 &26.84 &27.05& \textbf{27.93}\\
 \hline
 runway22 & x2 &27.24 &30.11 &30.39&31.19&33.38&33.68& \textbf{34.25}\\
& x3 &24.16&27.13&28.07&28.59&29.46&29.96 &\textbf{31.01}\\
& x4 & 23.82&23.17&24.81&25.32&26.49&26.95& \textbf{27.18}\\
 \hline
 Test Dataset & x2 &32.16&32.98&33.81&34.08&34.26&34.31& \textbf{35.21}\\
(Zebra)& x3 &28.32&29.13&30.49&31.14&31.79&30.76&\textbf{31.10}\\
& x4 &27.02&28.95&29.01&29.22&29.29&29.41 & \textbf{29.83}\\
 \hline
\end{tabular}
\end{table}

\begin{table}\label{Tab2}
  \caption{The SSIM value of HR reconstructed image compared to the latest methods}

  \begin{tabular}{ccccccccc}
   \hline
Image & Scale & Bicubic & SRCNN& FSRCNN & ESPCN& VDSR& SRGAN& \textbf{Proposed}\\
& & & \cite{dong01} & \cite{dong02}&\cite{shi02}&\cite{kim02}&\cite{ledig01}&\textbf{(TWIST-GAN)}\\
    \midrule
    airplane41 & x2 & 0.9156 & 0.9485 & 0.9521 & 0.9501 & 0.9537 & 0.9557 & \textbf{0.9607}\\
& x3 & 0.8346 & 0.8764& 0.8862 & 0.8831 & 0.8877 & 0.8979&\textbf{0.9117}\\
& x4 & 0.7672 & 0.8026  & 0.8165 & 0.8128 & 0.8189 & 0.8155&\textbf{0.8607}\\
 \hline
  airplane47 & x2 & 0.9388 &0.9596 &0.9604 &0.9589 &0.9616 &0.9625& \textbf{0.9634}\\
& x3 & 0.8616 &0.8873&0.8911&0.8877&0.8961 &0.8977&\textbf{0.8987}\\
& x4 & 0.7992 &0.8133 & 0.8241&0.8185&0.8294&0.8319& \textbf{0.8337}\\
 \hline
 airplane90 & x2 & 0.9142 &0.9468 &0.9512&0.9505&0.9529&0.9553 & \textbf{0.9587}\\
& x3 & 0.8327 &0.8735 &0.8823 &0.8811&0.8839&0.8956&\textbf{0.8981}\\
& x4 & 0.7672 &0.8009& 0.8119&0.8110 &0.8231 &0.8353& \textbf{0.8391}\\
 \hline
 runway22 & x2 &0.8786 &0.9179 &0.9231&0.9188&0.9310&0.9354& \textbf{0.9401}\\
& x3 &0.7849&0.8254&0.8472&0.8352&0.8584&0.8631 &\textbf{0.9012}\\
& x4 & 0.7132&0.7321&0.7525& 0.7218 &0.7618&0.7708& \textbf{0.8417}\\
 \hline
 Test Dataset & x2 &0.9046&0.9322&0.9387&0.9373&0.9450&0.9527& \textbf{0.9617}\\
(Zebra)& x3 &0.8126&0.8678&0.8705&0.8695& 0.8789&0.8878&\textbf{0.9123}\\
& x4 &0.7576&0.7916&0.8226&0.8213&0.8294&0.8514 & \textbf{0.8632}\\
 \hline
\end{tabular}
\end{table}

\begin{figure*}[ht]
\includegraphics[width=1\textwidth]{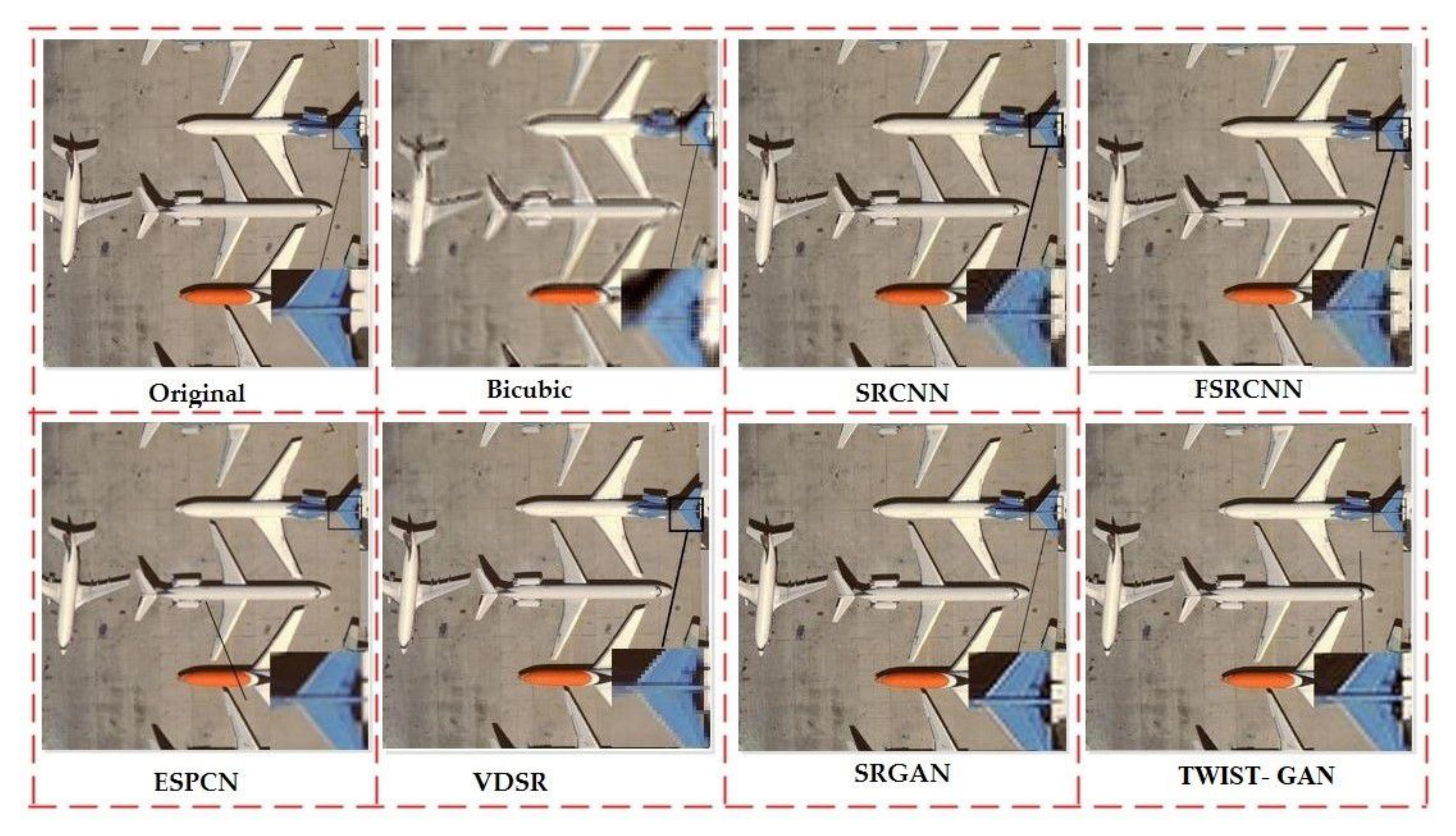}
\caption{Comparison of reconstructed HR images of "airplane 90.jpg" obtained from UCMerced LandUse dataset class "airplane" with 256x256 pixel images using different methods with a scale factor of x2.}
\centering\label{Fig8}
\end{figure*}

\begin{figure*}[ht]
\includegraphics[width=1\textwidth]{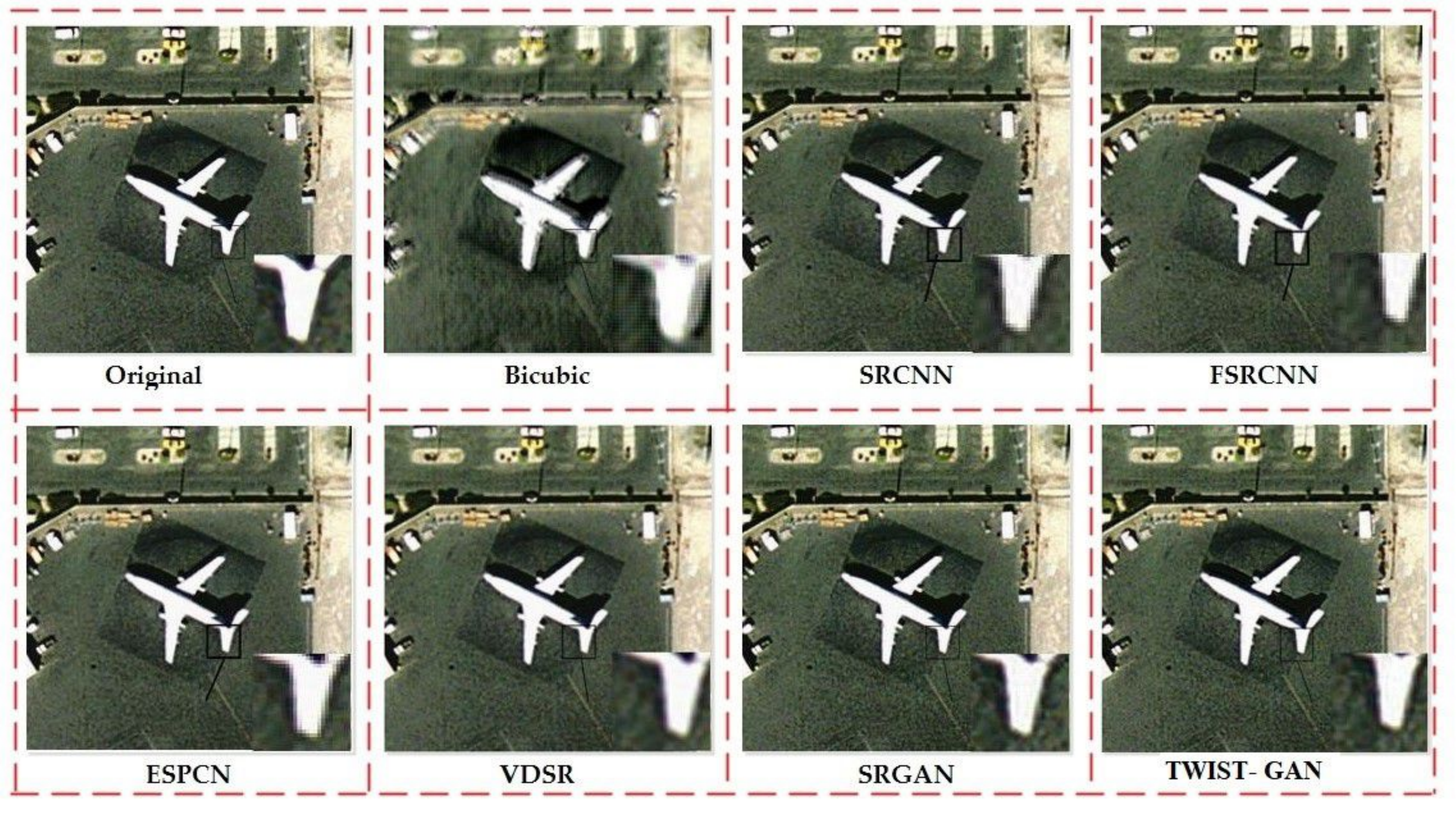}
\caption{Comparison of reconstructed HR images of "airplane 41.jpg" obtained from UCMerced LandUse dataset class "airplane" with 256x256 pixel images using different methods with a scale factor of x2.}
\centering\label{Fig9}
\end{figure*}

\begin{figure*}[ht]
\includegraphics[width=1\textwidth]{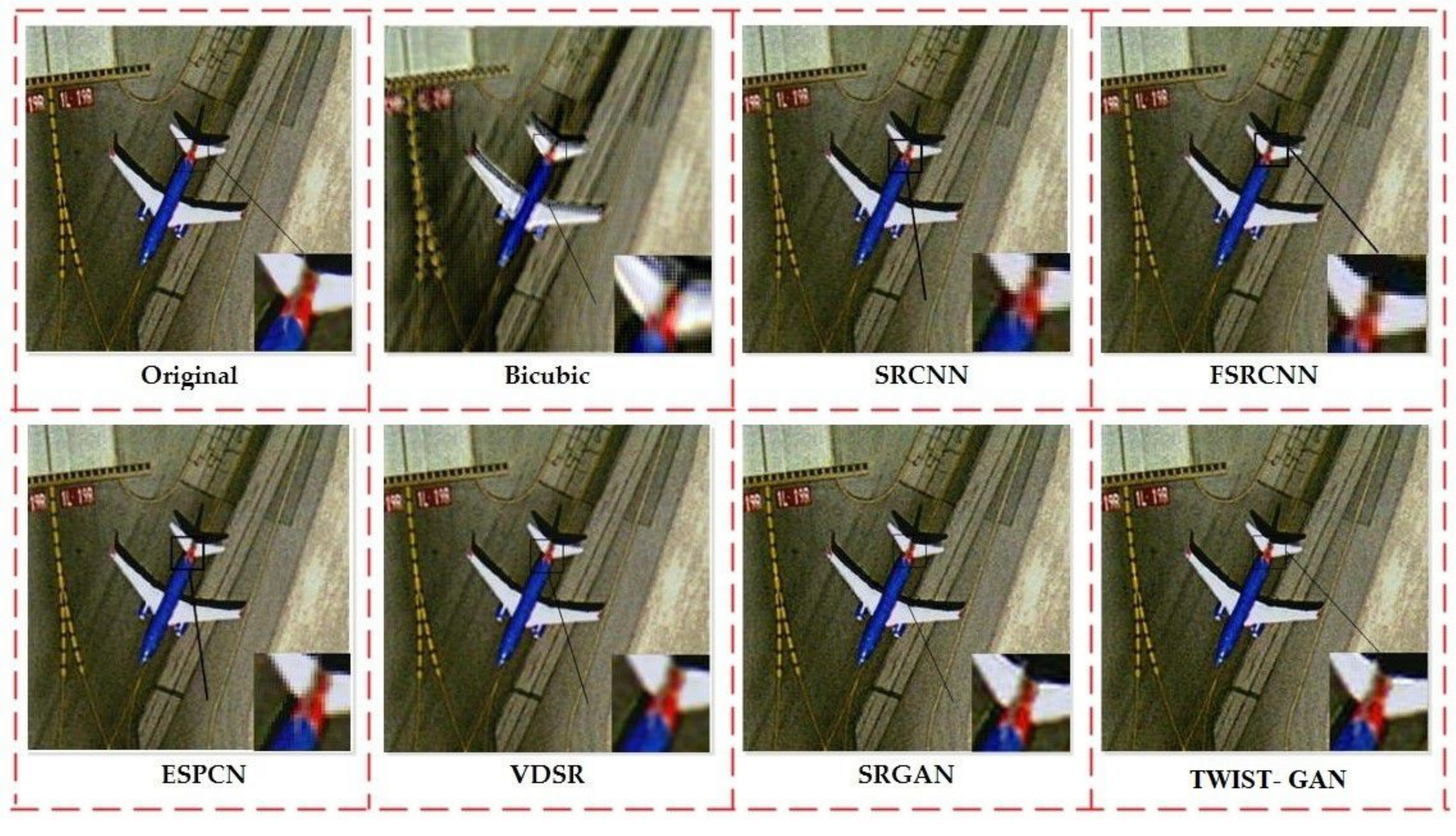}
\caption{Comparison of reconstructed HR images of "airplane 47.jpg" obtained from UCMerced LandUse dataset class "airplane" with 256x256 pixel images using different methods with a scale factor of x2.}
\centering\label{Fig10}
\end{figure*}

\begin{figure*}[ht]
\includegraphics[width=1\textwidth]{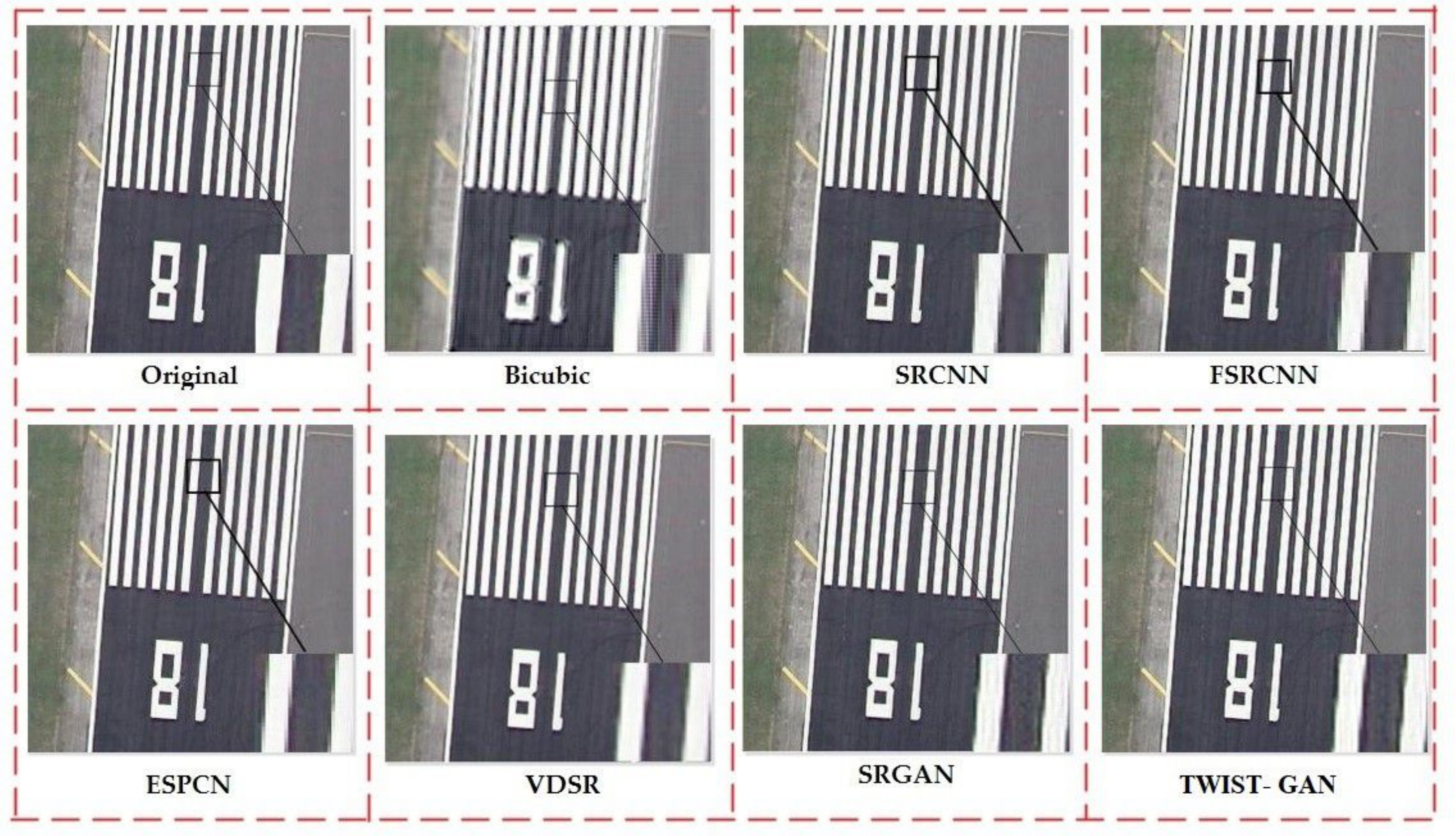}
\caption{Comparison of reconstructed HR images of "runway 22.jpg" obtained from UCMerced LandUse dataset class "runway" with 256x256 pixel images using different methods with a scale factor of x2.}
\centering\label{Fig11}
\end{figure*}

\begin{figure*}[ht]
\includegraphics[width=1\textwidth]{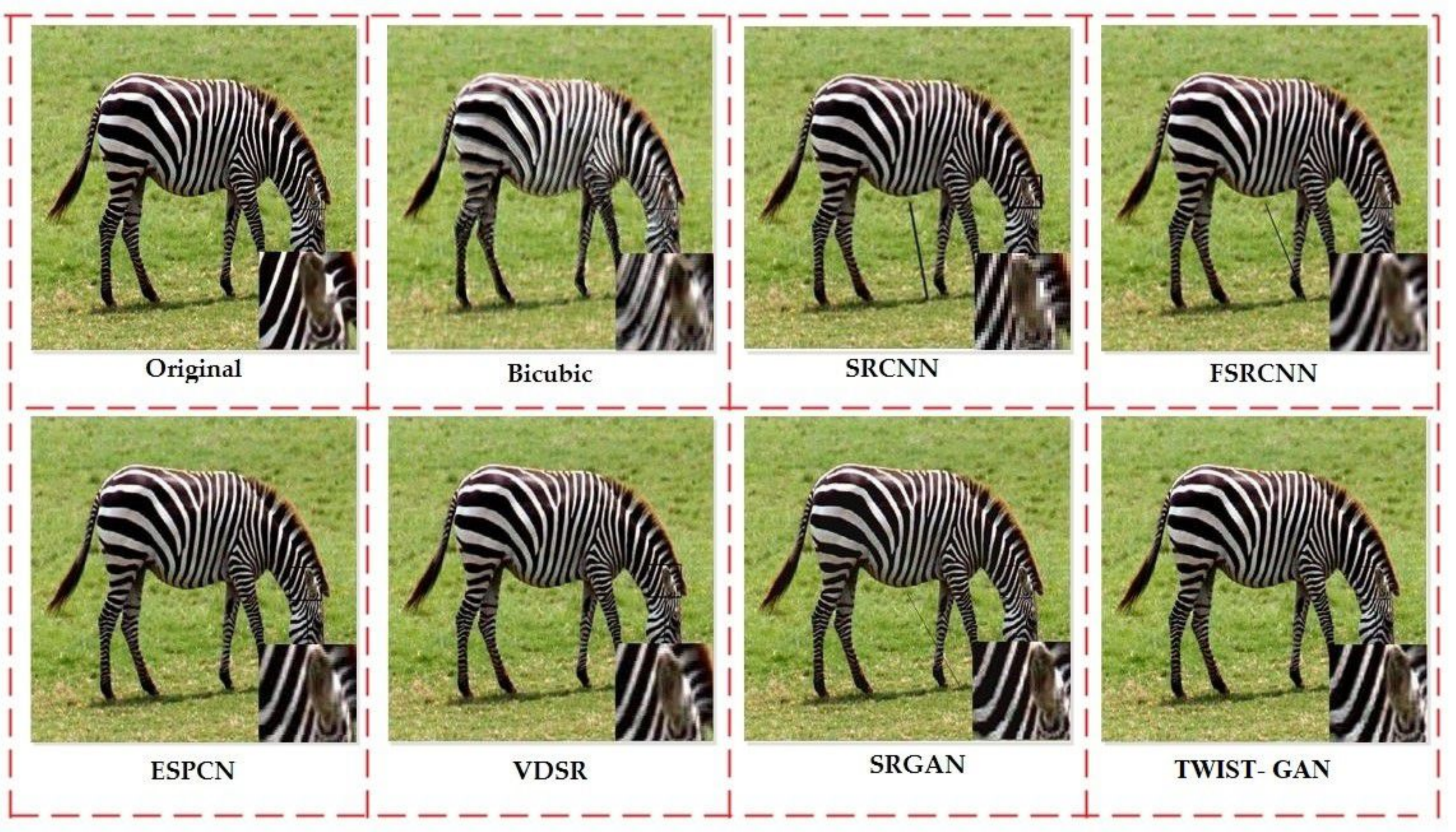}
\caption{Comparison of reconstructed HR images of "freeway 39.jpg" obtained from set14 dataset class "zebra" with resized 256x256 pixel images using different methods with a scale factor of x2.}
\centering\label{Fig12}
\end{figure*}

As shown in Figures. \ref{Fig8}, \ref{Fig9}, \ref{Fig10},  \ref{Fig11}, we evaluated our model on UCMerceed Landuse remote sensing dataset and also compared it with  multiple methods, SRCNN \city{dong01}, FSRCNN \city{dong02}, EPCN \city{shi01}, VDSR \city{kim01}, and SRGAN \city{lideg01}. We also verify our outcome with the test dataset from set14, shown in Figure. \ref{Fig12}. In visual as well as objective terms, the endorsed approach has been shown to perform better than all other strategies.We use FSIM and UIQ to evaluate the reconstructed image and analyze the details to evaluate the reconstructed image's accuracy more effectively. Table. 3 and Table. 4 demonstrate the effects of our test images on FSI Mand UIQ. The better the restored image's quality, the closer the value of FSIM and UIQ is to 1. We propose a wavelet and GAN-based super-resolution reconstruction algorithm to enhance the restored image and HR image generation's objective evaluation with a richer texture.
\begin{table}\label{Tab3}
  \caption{The FSIM value of HR reconstructed image compared to the latest methods}

  \begin{tabular}{ccccccccc}
   \hline
Image & Scale & Bicubic & SRCNN& FSRCNN & ESPCN& VDSR& SRGAN& \textbf{Proposed}\\
& & & \cite{dong01} & \cite{dong02}&\cite{shi02}&\cite{kim02}&\cite{ledig01}&\textbf{(TWIST-GAN)}\\
    \midrule
    airplane41 & x2 & 0.896 & 0.935 & 0.942 & 0.937 & 0.948 & 0.952 & \textbf{0.965}\\
& x3 & 0.814 & 0.878& 0.879 & 0.877 & 0.896 & 0.907&\textbf{0.917}\\
& x4 & 0.757 & 0.807  & 0.815 & 0.808 & 0.825 & 0.834&\textbf{0.852}\\
 \hline
  airplane47 & x2 & 0.902 &0.938 &0.943 &0.931 &0.946 &0.951& \textbf{0.954}\\
& x3 & 0.825 &0.884 &0.888 & 0.879 &0.893 & 0.906 &\textbf{0.913}\\
& x4 & 0.769 &0.819 & 0.826 & 0.817 &0.837 &0.838 & \textbf{0.847}\\
 \hline
 airplane90 & x2 & 0.889 &0.931 &0.940 &0.933&0.944&0.948 & \textbf{0.959}\\
& x3 & 0.809 &0.874 &0.875 &0.872 &0.893 &0.903&\textbf{0.912}\\
& x4 & 0.752 &0.803& 0.812&0.804 &0.821 &0.830& \textbf{0.848}\\
 \hline
 runway22 & x2 &0.881 &0.918 &0.924&0.917&0.928&0.927& \textbf{0.933}\\
& x3 &0.807&0.853&0.861&0.860&0.867&0.892 &\textbf{0.907}\\
& x4 & 0.735&0.799&0.807& 0.802 &0.817&0.821& \textbf{0.849}\\
 \hline
 Test Dataset & x2 &0.892&0.931&0.938&0.931&0.942&0.948& \textbf{0.951}\\
(Zebra)& x3 &0.812&0.874&0.882&0.872& 0.891&0.902&\textbf{0.912}\\
& x4 &0.754&0.802&0.819&0.797&0.821&0.829 & \textbf{0.847}\\
 \hline
\end{tabular}
\end{table}

\begin{table}\label{Tab4}
  \caption{The UIQ value of HR reconstructed image compared to the latest methods}

  \begin{tabular}{ccccccccc}
   \hline
Image & Scale & Bicubic & SRCNN& FSRCNN & ESPCN& VDSR& SRGAN& \textbf{Proposed}\\
& & & \cite{dong01} & \cite{dong02}&\cite{shi02}&\cite{kim02}&\cite{ledig01}&\textbf{(TWIST-GAN)}\\
    \midrule
    airplane41 & x2 & 0.889 & 0.928 & 0.932 & 0.931 & 0.938 & 0.948 & \textbf{0.957}\\
& x3 & 0.808 & 0.865& 0.867 & 0.866 & 0.893 & 0.902&\textbf{0.914}\\
& x4 & 0.749 & 0.798  & 0.806 & 0.802 & 0.819 & 0.830&\textbf{0.848}\\
 \hline
  airplane47 & x2 & 0.901 &0.929 &0.935 &0.931 &0.941 &0.948& \textbf{0.950}\\
& x3 & 0.818 &0.872 &0.881 & 0.875 &0.891 & 0.902 &\textbf{0.911}\\
& x4 & 0.755 &0.817 & 0.823 & 0.814 &0.833 &0.834 & \textbf{0.844}\\
 \hline
 airplane90 & x2 & 0.883 &0.926 &0.930 &0.928&0.937&0.939 & \textbf{0.943}\\
& x3 & 0.805 &0.861 &0.866 &0.865 &0.873 &0.896&\textbf{0.905}\\
& x4 & 0.748 &0.796& 0.804&0.797 &0.807 &0.826& \textbf{0.839}\\
 \hline
 runway22 & x2 &0.887 &0.916 &0.922&0.915&0.925&0.924& \textbf{0.929}\\
& x3 &0.805&0.851&0.860&0.858&0.864&0.890 &\textbf{0.904}\\
& x4 & 0.746&0.796&0.804& 0.799 &0.814&0.818& \textbf{0.843}\\
 \hline
 Test Dataset & x2 &0.888&0.917&0.925&0.916&0.927&0.926& \textbf{0.932}\\
(Zebra)& x3 &0.807&0.855&0.862&0.859& 0.866&0.893&\textbf{0.906}\\
& x4 &0.747&0.799&0.807&0.803&0.818&0.820 & \textbf{0.846}\\
 \hline
\end{tabular}
\end{table}

The run-time of all algorithms is illustrated in Table. 5. The running time of the conventional algorithm is longer than that of the deep learning algorithms. Bicubic shows the shortest time-consuming because it only has interpolation operations. SRCNN and VDS take a longer time between LR and HR image patch pairs. SRGAN and ESPCN models have a wide range of convolutional layers, so they all take a long time to train TWIST-GAN to reduce network depth by eliminating BN layers and not increasing network time complexity. Although FSRCNN performs better, it does not perform well in visual  quality,  such  as  PSNR  and  SSIM.  We  performed  experiments  with  the  same  hardware  setup,  and  carried  out  all algorithms: Intel Core i7-6700K C3.20 GHz CPU via an NVIDIA GTX1080 GPU 8 GB RAM. 

\begin{table}\label{Tab5}
  \caption{Total running time of different algorithms on UCMerced dataset.}

  \begin{tabular}{ccccccccc}
   \hline
Algorithms & Bicubic & SRCNN& FSRCNN & ESPCN& VDSR& SRGAN& \textbf{Proposed}\\
& &\cite{dong01} & \cite{dong02}&\cite{shi02}&\cite{kim02}&\cite{ledig01}&\textbf{(TWIST-GAN)}\\
    \midrule
    Running/Time & 10.269 & 15.144 & 12.025 & 14.301 & 17.442 &16.278 & \textbf{13.935}\\

 \hline
\end{tabular}
\end{table}

\subsection{Model Convergence}
\begin{figure}[ht]
\includegraphics[width=1\textwidth]{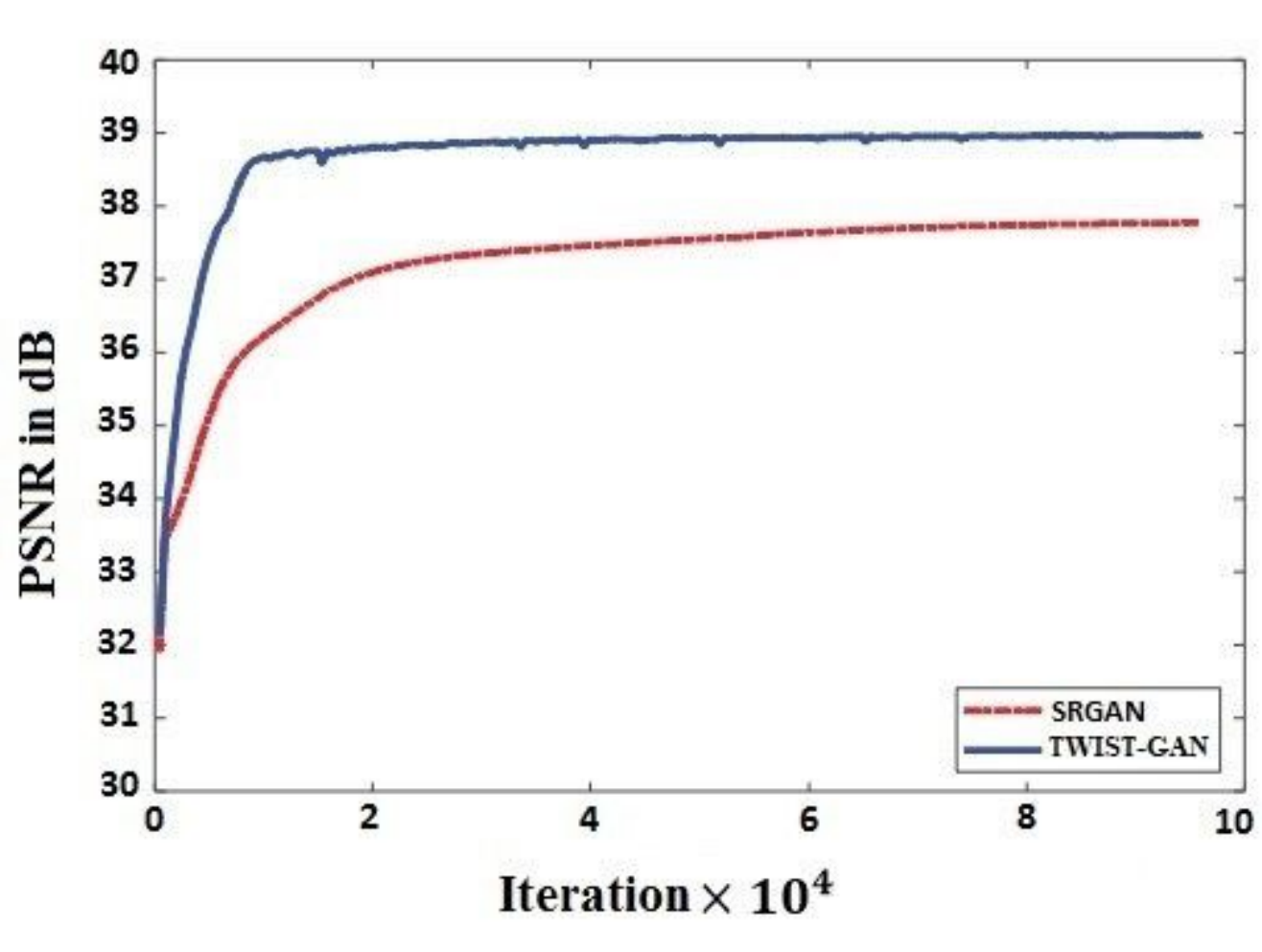}
\caption{Comparison of reconstructed HR images of "freeway 39.jpg" obtained from set14 dataset class "zebra" with resized 256x256 pixel images using different methods with a scale factor of x2.}
\centering\label{Fig13}
\end{figure}

Our TWIST-GAN improves SRGAN by training multiple networks to characterize wavelets. We compare their training convergence speeds. The experiments are carried out in the same network setup and computing environment. The curves of  convergence  regarding  PSNR  are  shown  in  Figure.\ref{Fig13}.  Within  8  to 104of  the  SRGAN is  converged  to  our  TWIST-GAN optima with better accuracy of 4 to 104iterations. TWIST-GAN restores the entire aerial image straight away, and transfers the whole aerial image.

\section{CONCLUSION}
We propose a method to provide the super-resolution performance of remote sensing images by wavelet transformation and transferred generative adversarial network. The computational cost of our approach is minimized, and the similarity of spatio-temporal remote-sensing data is increased by preprocessing spatio-temporal remote-sensing data. The method is validated by several sets of experiments conducted on common remote sensing data. We first trained our network on the high-resolution DIV2K dataset, and then applied transfer learning techniques to train and test the remote sensing images on the UC Merceed dataset. The objective and subjective results demonstrate the significance of our proposed approach. The experimental results suggest that the proposed method can effectively develop the objective assessment limitations, and the  reconstructed images  are  more  realistic  images  with  more  texture  details.  This  demonstrates  that  our proposed algorithm got the advantage of wavelet packet transformation and Transfer GAN in spatio-temporal remote sensing image super-resolution. Although the proposed method is effective, it can still be improved by applying other techniques in future work, such as multi-resolution discrete wavelet transform and dual-tree complex wavelet transform. High-frequency data from natural image data sets can be extracted using these techniques.

\section{Acknowledgments}
This work was supported in part by the Key Research Program of Frontier Sciences, CAS, and Grant number ZDBS-LY-DQC016, Beijing Natural Science Foundation under Grant No. 4212030, Beijing Nova Program of Science and Technology under Grant No. Z191100001119090, Natural Science Foundation of China under Grant No. 61836013 and, Youth Innovation Promotion Association CAS.

\bibliographystyle{ACM-Reference-Format}
\bibliography{sample-base}

\end{document}